# Sensitivity of Spin-Torque Diodes for Frequency-Tunable Resonant Microwave Detection

C. Wang[1], Y.-T. Cui[1], J. Z. Sun[2], J. A. Katine[3], R. A. Buhrman[1], and D. C. Ralph[1]

[1]*Cornell University, Ithaca, New York 14853, USA*
[2]*IBM T. J. Watson Research Center, Yorktown Heights, New York 10598, USA*
[3]*Hitachi Global Storage Technologies, San Jose Res. Ctr., San Jose, CA 95135 USA*

We calculate the efficiency with which magnetic tunnel junctions can be used as resonant detectors of incident microwave radiation via the spin-torque diode effect. The expression we derive is in good agreement with the sensitivities we measure for MgO-based magnetic tunnel junctions with an extended (unpatterned) magnetic pinned layer. However, the measured sensitivities are reduced below our estimate for a second set of devices in which the pinned layer is a patterned synthetic antiferromagnet (SAF). We suggest that this reduction may be due to an undesirable coupling between the magnetic free layer and one of the magnetic layers within the etched SAF. Our calculations suggest that optimized tunnel junctions should achieve sensitivities for resonant detection exceeding 10,000 mV/mW.

The spin-transfer-torque effect[1,2,3] in magnetic tunnel junctions (MTJs) is under development for several types of applications,[4] including the switching of magnetic elements in nonvolatile magnetic random access memory[5,6,7] and the excitation of steady-state magnetic precession to produce nanometer-scale microwave sources.[8,9,10,11,12] MTJs also have the capability to be used as frequency-tunable resonant microwave detectors, via a process dubbed spin-torque diode detection[13] or spin-torque-driven ferromagnetic resonance (ST-FMR).[14,15,16] In this process, when a microwave signal with frequency close to the natural ferromagnetic resonance frequency of one of the electrodes of a magnetic tunnel junction is incident onto the device, the oscillating tunnel current that it induces can excite magnetic precession via spin transfer. The resistance oscillations that result from this precession mix with the oscillating current to produce an easily measurable DC voltage component across the tunnel junction. This effect has recently been used to make quantitative measurements of spin transfer torque vector in magnetic tunnel junctions, along with its bias dependence.[17,18,19] Here we use a similar analysis to discuss how to maximize the microwave detector sensitivity of a tunnel junction, which we define as $\varepsilon = \langle V_{res} \rangle / P_{inc}$, where $\langle V_{res} \rangle$ is the resonant part of the measured DC voltage (above a non-resonant background) and $P_{inc}$ is the incident microwave power. This sensitivity is of course only one of several parameters important for applications (e.g., background noise, dynamic range, and speed), but achieving a competitive detector sensitivity is a first prerequisite for evaluating whether more detailed studies are warranted.

We note first that the detector sensitivity can vary as a function of a DC bias applied across the tunnel junction.[19] However, in the devices measured thus far, we find that the maximum detection sensitivity is within 10% of the value at zero bias (see Fig. 1). For the purposes of estimating the typical detector sensitivity, we will therefore limit our discussion to the simple case of zero applied bias. The fall-off in sensitivity at large biases can be explained by a reduction in the tunnel magnetoresistance (TMR) as a function of bias, which decreases the size of the resistance oscillations contributing to the mixing signal. The maximum sensitivity is not found exactly at zero bias because there are several other mechanisms by which an applied bias can either enhance or suppress the microwave sensitivity. For example, the strength of the spin torque is bias dependent.[17,18,19] An incident microwave signal can also change the time-averaged junction resistance, which in the presence of a non-zero bias gives a contribution $I_{DC}\Delta R$ to the DC voltage on resonance in addition to the mixing contribution noted above.[17,19] Furthermore, for the correct sign of bias one can decrease the effective magnetic damping, thereby making the resonant detection more efficient. If future improvements in magnetic-tunnel-junction technology allow for a TMR that does not decrease strongly as a function of bias, it may be possible to take advantage of these other mechanisms by using a non-zero bias to improve the sensitivity beyond the value we estimate in this paper. The influence of these mechanisms on the detection sensitivity can be calculated using the methods described in the appendix of ref. [19].

For our estimate of the sensitivity, we consider a magnetic tunnel junction in which the magnetization of one electrode is pinned (e.g., by exchange bias to an antiferromagnetic layer) so that only a single magnetic layer (the "free layer") can undergo magnetic precession in response to the microwave signal. We assume that the volume of the free layer is sufficiently small that it can be modeled as a single macrospin, and that its magnetic orientation, $\hat{m}$, is initially oriented at an angle $\theta$ from the magnetization of the pinned layer, $\hat{M}_{pin}$. We define the $\hat{z}$ direction as parallel to the initial orientation of $\hat{m}$, the $\hat{x}$ axis as along $\hat{M}_{pin} \times \hat{z}$, and the $\hat{y}$ axis in the plane defined by $\hat{m}$ and $\hat{M}_{pin}$ such that $\hat{M}_{pin} \cdot \hat{y} > 0$. For mathematical simplicity we assume that the plane defined by $\hat{m}$ and $\hat{M}_{pin}$ is parallel to one of the symmetry planes of the magnetic anisotropy tensor for the free layer,[15,16] as is the case in all existing experiments. We calculate the magnetic dynamics by means of the Landau-Lifshitz-Gilbert-Slonczewski (LLGS) equation[1]

$$\frac{d\hat{m}}{dt} = -\gamma \hat{m} \times \vec{H}_{eff} + \alpha \hat{m} \times \frac{d\hat{m}}{dt} - \gamma \frac{\tau_{\parallel}}{M_S tA} \hat{y} - \gamma \frac{\tau_{\perp}}{M_S tA} \hat{x}. \qquad (1)$$

Here $\gamma = 2\mu_B / \hbar = 1.76 \times 10^7 \text{ G}^{-1}\text{s}^{-1}$ is the absolute value of the gyromagnetic ratio, $\mu_B$ is the Bohr magneton, $\vec{H}_{eff}$ accounts for total effective field acting on the precessing layer including both the external field ($H$) and demagnetization terms, $\alpha$ is the phenomenological Gilbert damping constant, $M_S$ is magnetic moment per unit volume of the free layer, $t$ its thickness and $A$ its area, $\tau_\parallel(I,\theta)$ and $\tau_\perp(I,\theta)$ are the in-plane and out-of-plane components of the spin-transfer torque, and $I$ is the current. We neglect any effects of the Oersted field due to the current.

If a microwave-frequency current $I_{RF}$ passes through the tunnel junction, the solution of the LLGS equation yields an expression for the resonant DC voltage signal due to the mixing between the spin-torque-driven magnetic precession and $I_{RF}$ (to lowest order in $I_{RF}$):[19]

$$\langle V_{res}\rangle = \frac{dR}{d\theta}\frac{\mu_B}{2\hbar(M_S tA)\sigma}I_{RF}^2\left[\frac{d\tau_\parallel}{dI}S(\omega) - \frac{d\tau_\perp}{dI}\frac{\gamma(H_z + 4\pi M_{eff})}{\omega_m}A(\omega)\right], \quad (2)$$

where $R$ is the differential resistance of the tunnel junction, $H_z$ is the $\hat{z}$ component of the external magnetic field, $4\pi M_{eff}$ is the effective out-of-plane anisotropy, and $S(\omega) = [1 + (\omega - \omega_m)^2/\sigma^2]^{-1}$ and $A(\omega) = [(\omega - \omega_m)/\sigma]S(\omega)$ are symmetric and antisymmetric components of the resonance lineshape as a function of frequency $\omega$, with $\omega_m$ the resonance frequency and $\sigma$ the linewidth. (We have neglected a small contribution from the within-plane anisotropy.) If the microwave signal is incident onto the tunnel junction from a transmission line with impedance $Z_0$ ($Z_0$ = 50 $\Omega$ for our apparatus), after taking into account the impedance mismatch with the junction the incident power can be related to the microwave current in the junction as[20]

$$P_{inc} = \frac{1}{2Z_0}\left[\frac{R+Z_0}{2}\right]^2 I_{RF}^2. \quad (3)$$

Therefore the overall sensitivity for the detected signal is

$$\varepsilon \equiv \langle V_{res}\rangle/P_{inc} = \frac{dR}{d\theta}\frac{4\mu_B}{\hbar(M_S tA)\sigma}\frac{Z_0}{(R+Z_0)^2}\left[\frac{d\tau_\parallel}{dI}S(\omega) - \frac{d\tau_\perp}{dI}\frac{\gamma(H_z + 4\pi M_{eff})}{\omega_m}A(\omega)\right]. \quad (4)$$

This expression can be simplified further. In the case we are considering, for zero applied bias on the tunnel junction, it is predicted theoretically[21,3] and observed experimentally[17,18] for a symmetric tunnel junction (with both magnetic electrodes made from the same material) that $d\tau_\perp/dI = 0$. Therefore, for this case the second term in the brackets of Eq. (4) is zero and the lineshape should be symmetric in frequency, with the maximum detector sensitivity occurring at the resonance frequency, for which $S(\omega_m) = 1$. (The cases of symmetric junctions with non-zero bias or asymmetric junctions may both be more complicated.) The angular dependence of the zero-bias tunnel junction conductance is expected to be purely sinusoidal,[22]

$$\frac{1}{R} = \frac{1}{2}\left(\frac{1}{R_P} + \frac{1}{R_{AP}}\right) + \frac{1}{2}\left(\frac{1}{R_P} - \frac{1}{R_{AP}}\right)\cos\theta, \tag{5}$$

so that

$$\frac{dR}{d\theta} = \frac{R^2}{2R_P R_{AP}}(R_{AP} - R_P)\sin\theta. \tag{6}$$

Finally, for a symmetric tunnel junction the in-plane spin torque is predicted to have the magnitude[23,24]

$$\frac{d\tau_\parallel}{dI} = \frac{\hbar}{4e}\frac{2P}{1+P^2}\frac{R}{R_P}\sin\theta, \tag{7}$$

where $P^2 = (R_{AP} - R_P)/(R_{AP} + R_P)$. ST-FMR measurements[17] have found good quantitative agreement with Eq. (7). Incorporating these values into Eq. (4), we reach an expression for the maximum detector sensitivity of a symmetric magnetic tunnel junction for zero applied bias:

$$\varepsilon = \frac{R_{AP} - R_P}{R_P}\frac{\mu_B}{2e(M_S tA)\sigma}\frac{RZ_0}{(R+Z_0)^2}\frac{2P}{1+P^2}\frac{R^2}{R_P R_{AP}}\sin^2\theta. \tag{8}$$

We have tested this estimate using two batches of MgO-based magnetic tunnel junctions. The first batch, with an average resistance-area product for parallel magnetic alignment of $RA = 12$ Ω-μm$^2$, had the layer structure (in nm) Ta(5)/Cu(20)/Ta(3)/Cu(20)/PtMn(15)/Co$_{70}$Fe$_{30}$(2.5)/Ru(0.85)/Co$_{60}$Fe$_{20}$B$_{20}$(3)/MgO(1.25)/Co$_{60}$Fe$_{20}$B$_{20}$(2.5)/Ta(5)/Ru(7) deposited on an oxidized silicon wafer. The top ("free") magnetic layer of these samples was etched to be a rounded rectangle, with nominal dimensions either 50 × 100 nm$^2$ or 50 × 150 nm$^2$, and had a saturation magnetization per unit area measured to be $M_S t = 2.75 \times 10^{-4}$ e.m.u./cm$^2$. The milling process used to define the nanopillar was stopped at the tunnel barrier, leaving the bottom magnetic electrode extended. The exchange bias for the bottom electrode was parallel to the magnetic easy axis of the top layer. The second batch of samples, with $RA = 1.5$ $\Omega$-$\mu m^2$, had the layer structure (in nm) Ta(3)/CuN(41.8)/Ta(3)/CuN(41.8)/Ta(3)/Ru(3.1)/IrMn(6.1)/CoFe(1.8)/Ru/CoFeB(2)/ MgO$_x$/CoFe(0.5)/CoFeB(3.4)/Ru(6)/Ta(3)/Ru(4). These samples were etched to 90 nm diameter circles, with the etch extending through both the CoFe/CoFeB composite free layer and the IrMn/CoFe/Ru/CoFeB exchange-biased synthetic antiferromagnetic (SAF) pinned layer. For these free layers, $M_S t = 3.2 \times 10^{-4}$ e.m.u./cm$^2$. We measured more than 5 samples from each batch, and found good consistency for each sample type, with microwave detection sensitivities varying within a range of 20%. Here we will report results from 3 samples, whose parameters are listed in Table 1.

The measurements were performed at room temperature by contacting each sample to a 50 $\Omega$ semirigid coaxial cable via a high-frequency probe station. A magnetic field was applied within the sample plane at various directions to control the initial offset angle $\theta$ between the magnetic orientations of the two electrodes. We determined $\theta$ from the magnetoresistance [Eq. (5)], measured in situ using a lock-in amplifier. To measure the microwave detection sensitivity, we swept the frequency of an incident microwave signal while keeping the power, $P_{inc}$, constant. The incident microwaves were chopped as a function of time, and the resonant DC detector mixing voltage was measured using a second lock-in amplifier having a large input impedance and connected to the sample via a bias tee. The incident microwave power was calibrated by using a non-resonant background voltage in the mixing signal arising from the nonlinear tunnel junction resistance to determine $I_{RF}$ within the sample, and then using Eq. (3) to evaluate $P_{inc}$.[17] The applied power was kept small enough that the magnetic precession angle was always < 0.5°, within the linear-response regime. The dominant uncertainty in comparing the measured and predicted values of the detection sensitivity comes from estimating the sample area using scanning electron microscope pictures. The true areas are less than the nominal values listed in Table 1 because the pillars do not have perpendicular sidewalls. We estimate that our values of the true areas (and hence $M_S t A$) are accurate to ±15%.

The measured mixing voltage as a function of frequency is shown in Fig. 2(a) for sample #1 for various offset angles, $\theta$. Both the magnitude and angle of the applied magnetic field were varied (as noted in the figure caption) to access a wide range of $\theta$. In each spectrum in Fig. 2(a) we observe only a single large resonant peak, with a peak shape that to a good approximation is a symmetric Lorentzian, as predicted. The measured microwave sensitivities corresponding to the maxima of these resonance curves are listed in Table 2. For sample #1, the measurements show excellent quantitative agreement with the values predicted by Eq. (8), within the estimated uncertainty of ~15% associated with the determination of the sample area. The detector sensitivity is predicted to be maximal near 90° (to be precise, at $\theta = \arccos\left[\left(-1+\sqrt{1-P^4}\right)/P^2\right]$ in the limit $R \gg Z_0$, or slightly greater than 90°), since this maximizes both the spin torque and the response of the resistance to changes in $\theta$. The angular dependence of the data is in good agreement. Sample #2, with the same $RA$ product but a larger area than sample #1, exhibits very similar maximum detection sensitivities (see Table 2). This is also in agreement with Eq. (8). In the limit that the tunnel junction resistance $R$ is much greater then $Z_0$ (50 $\Omega$), the sensitivity is predicted to depend on $R$ and $A$ only through their product $RA$, which is independent of the area $A$. Physically, the explanation for the area-independent sensitivity is that the better impedance matching provided by the lower resistance of sample #2 is offset by its larger total magnetic moment. The excellent quantitative agreement between our estimate (Eq. (8)) and the sensitivities measured for samples #1 and #2 give us confidence in the reliability of our estimate.

The measured mixing voltage as a function of frequency for a lower-$RA$ sample (sample #3) is shown in Fig. 2(b) for various offset angles. Here we observe two peaks in each spectrum, with a significant degree of asymmetry in the peak shapes as a function of frequency, in contrast to the other samples which exhibited only a single large peak symmetric in frequency. We attribute this difference to the fact that the pinned magnetic electrode in samples #1 and #2 is left as an unetched extended film, while the pinned electrode in sample #3 is etched. This etching leaves the upper CoFeB layer within the CoFe/Ru/CoFeB SAF in sample #3 free to precess in response to a spin torque, giving a second resonant mode in addition to the free layer. Magnetic coupling between these two modes can induce the asymmetric peak shapes seen in Fig. 2(b).[25] We use the larger, lower-frequency resonance peak to determine the resonant detector response $\langle V_{res} \rangle$ (listed in Table 2).

We expect that the sensitivity of sample #3 should be greater than for the other samples, due primarily to its lower *RA* product, which enables better impedance matching to the 50 Ω input waveguide. Indeed, sample #3 exhibits the largest peak sensitivity we have observed to date, 54 mV/mW for $\theta$ near 100°. However, the measured sensitivities for sample #3 do not agree with Eq. (8), in all cases being approximately 30-35% below the predicted values. One potential explanation for this difference could in principle be a weaker spin torque in these low-*RA* samples than expected from Eq. (7). We suggest that a more likely cause is coupling between the free layer and the top CoFeB layer within the CoFe/Ru/CoFeB SAF "pinned layer". Identifying the larger, lower-frequency resonance peaks in Fig. 2(b) as the acoustic mode in which these two layers precess with the same phase, coupled motion of these layers would reduce the detector sensitivity both because the relative excitation angle would be reduced, thereby reducing the size of the resistance oscillation, and because the extra volume of precessing magnetic material would produce an effectively larger value of $M_S$ in Eq. (8). To achieve the goal of maximizing the detector sensitivity in spin-torque diodes, we therefore suggest that the pinned layer should be left unetched so that exchange coupling to the extended film can suppress any tendency of the pinned layer to precess, thereby leaving the free layer without undesired coupling to other magnetic modes.

None of the devices that we have studied thus far were designed with optimization of the detector sensitivity in mind. Equation (8) suggests several strategies for significant improvement. For best sensitivity, one should maximize the TMR while minimizing the free layer magnetization $M_S$ and the free-layer thickness. In particular, the free-layer thicknesses can be reduced significantly below the values used for the samples in this paper (2.5 nm and 3.9 nm). For a given free layer volume, Eq. (8) predicts that the maximum sensitivity should be achieved by tuning the thickness of the tunnel barrier (or equivalently, the *RA* product) so as to match the tunnel junction resistance to the impedance of the waveguide ($Z_0$, typically 50 Ω) from which the microwaves are incident. However, for a given barrier thickness (*i.e.*, a given *RA* product) there is no advantage to increasing the tunnel junction area in order to improve the impedance match. We have already noted above in comparing samples #1 and #2 that for the case $R \gg Z_0$, Eq. (8) predicts that the sensitivity should be independent of the junction area for a given value of *RA*. For the case that *R* is comparable to $Z_0$, for a given barrier thickness the sensitivity should be optimized by always minimizing the junction area, even if this increases the impedance mismatch.

According to Eq. (8), the microwave sensitivity is also optimized by minimizing the resonance linewidth, $\sigma$. For the usual case that both the applied magnetic field and the initial orientation of the free layer moment are in the sample plane, the solution of the LLGS equation predicts[17,19] that

$$\sigma_{\text{in plane}} \approx \alpha\gamma(H_z + 2\pi M_{\text{eff}}), \tag{9}$$

where $4\pi M_{\text{eff}}$ is the effective demagnetization field perpendicular to the plane, and we have neglected a small contribution from the within-plane anisotropy. The measured linewidths in our samples agree with this expression using the parameters $\alpha = 0.010 \pm 0.002$ and $4\pi M_{\text{eff}} = 11 \pm 1$ kOe for samples #1 and #2, and $\alpha = 0.014 \pm 0.002$ and $4\pi M_{\text{eff}} = 13 \pm 1$ kOe for sample #3.[19] For the case that *H* is small relative to $4\pi M_{\text{eff}}$, in this in-plane orientation the microwave sensitivity should be relatively insensitive to the applied magnetic field, except through the effect of *H* on the offset angle $\theta$, and the detection sensitivity should scale inversely with $4\pi M_{\text{eff}}$. Consequently, efforts[26] to use interface anisotropy effects to reduce $4\pi M_{\text{eff}}$ well below the value $4\pi M_S$ could provide dramatic improvements to the detector sensitivity, as long as the Gilbert damping, $\alpha$, remains small in the process. The linewidth could also be decreased by designing the tunnel junction so that the free layer magnetization points out of plane, either through the use of materials giving a perpendicular anisotropy[27] or by using a perpendicularly applied magnetic field.[14] In these limits the form of our estimate for the sensitivity (Eq. (8)) is unchanged, but the LLGS prediction for the linewidth is[14]

$$\sigma_{\perp} = \alpha\gamma(H_z \pm 4\pi M_{\text{eff}}), \tag{10}$$

where the plus sign corresponds to an anisotropy favoring a perpendicular magnetization and the minus sign corresponds to an easy-plane anisotropy.

By extrapolating somewhat the current state of the art in MgO-based tunnel junctions, we estimate that one should be able to achieve the following parameters for a simple tunnel junction with an easy-plane magnetic anisotropy: $t = 1.0$ nm, $RA = 1$ $\Omega$-$\mu m^2$, $(R_{AP} - R_P)/R_P = 100\%$, $\alpha = 0.01$, $4\pi M_S = 10$ kOe, $4\pi M_{eff} = 1$ kOe,[26] and $A = \pi(50$ nm$)^2/4$. Using these parameters, Eq. (8) predicts an achievable sensitivity for resonant microwave detection of 10,300 mV/mW. In comparison, the zero-bias Schottky diode detectors used commonly for microwave detection at room temperature have sensitivities an order of magnitude less, 500-1000 mV/mW,[28] and are not frequency-tunable. Of course in judging the suitability for applications one should consider the signal-to-noise ratio and other figures of merit, not just the sensitivity. The noise limit in diode detectors is likely to be governed by thermal fluctuations in the orientation of the precessing magnetization, which increase with decreasing volume for the magnetic free layer. This effect may ultimately limit the effectiveness of increasing the detector sensitivity by simply decreasing the sample size. Techniques for calculating the effects of thermal fluctuations in spin-torque devices are under development,[29] but these techniques have not yet been applied to the diode-detector problem.

The detector sensitivity might be improved even more by making devices with two magnetic pinned layers rather than just one (one pinned layer on either side of the free layer). This geometry has been shown to give a factor of 2 or more increase in the strength of the spin torque on the free layer per unit current.[30,31,32]

In summary, we have derived an estimate of the sensitivity of resonant microwave detection by magnetic tunnel junctions used as spin-torque diodes. Our estimate is in excellent quantitative agreement with the measured sensitivities for $RA = 12$ $\Omega$-$\mu m^2$ MgO-based magnetic tunnel junctions in which the pinned magnetic electrode is left unetched. The measured sensitivity is decreased below our estimates in $RA = 1.5$ $\Omega$-$\mu m^2$ tunnel junctions in which the pinned layer consists of a synthetic antiferromagnet which is etched so that the top magnetic layer of the SAF is able to precess. We suggest that coupled precession of the free layer and this top layer of the SAF may be the cause of the reduced detector sensitivity. The maximum sensitivity that we have observed experimentally to date in our non-optimized tunnel junctions is 54 mV/mW. Our estimate for the detector sensitivity (Eq. (8)) suggests that device optimization should be able to improve the sensitivity to greater than 10,000 mV/mW.

We thank Y. Nagamine, D. D. Djayaprawira, N. Watanabe, and K. Tsunekawa of Canon ANELVA Corp. and D. Mauri of Hitachi Global Storage Technologies (now at Western Digital Corp.) for providing the junction thin film stacks that we used to fabricate the tunnel junctions. Cornell acknowledges support from DARPA, ARO, NSF (DMR-0605742), and the NSF/NSEC program through the Cornell Center for Nanoscale Systems. We also acknowledge NSF support through use of the Cornell Nanofabrication Facility/NNIN and the Cornell Center for Materials Research facilities.

| Sample # | nominal cross section | RA value | $R_P$ | $R_{AP}$ | $P$ | $M_s tA$ |
|---|---|---|---|---|---|---|
| 1 | 50 × 100 nm$^2$ | 12 Ω-μm$^2$ | 3.9 kΩ | 10.2 kΩ | 0.67 | 1.1 × 10$^{-14}$ emu |
| 2 | 50 × 150 nm$^2$ | 12 Ω-μm$^2$ | 2.21 kΩ | 5.89 kΩ | 0.67 | 1.6 × 10$^{-14}$ emu |
| 3 | 90 × 90 nm$^2$ | 1.5 Ω-μm$^2$ | 279 Ω | 537 Ω | 0.56 | 1.8 × 10$^{-14}$ emu |

**Table 1**. Sample parameters.

| Sample # | $\theta$ (deg.) | R (k$\Omega$) | $\sigma/(2\pi)$ (GHz) | measured $\varepsilon$ (mV/mW) | predicted $\varepsilon$ (mV/mW) |
|---|---|---|---|---|---|
| 1 | 49 | 4.36 | 0.197 | 12.3 | 10 |
| | 64 | 4.71 | 0.159 | 18.5 | 19 |
| | 95 | 5.86 | 0.174 | 24.6 | 27 |
| | 120 | 7.23 | 0.179 | 22.8 | 24 |
| | 142 | 8.65 | 0.188 | 15.7 | 14 |
| | 160 | 9.65 | 0.150 | 8.7 | 6 |
| 2 | 42 | 2.41 | 0.179 | 7.4 | 9 |
| | 96 | 3.37 | 0.188 | 25.0 | 28 |
| | 153 | 5.42 | 0.167 | 11.0 | 11 |
| 3 | 56 | 0.314 | 0.327 | 21.9 | 33 |
| | 87 | 0.363 | 0.286 | 42.6 | 65 |
| | 103 | 0.403 | 0.270 | 54.2 | 76 |

**Table 2**. Results of the detector diode sensitivity measurements, with a comparison to the sensitivity predicted by Eq. (8), calculated using the measured values of $\sigma$, $R$, $R_P$, and $R_{AP}$.

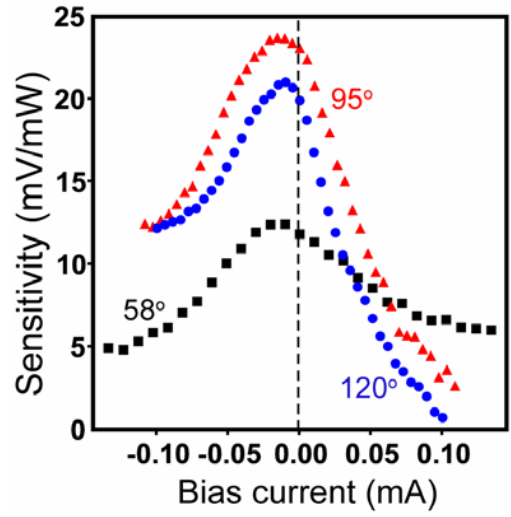

**Fig. 1**. (Color online) Bias dependence of the detector-diode sensitivity for sample #1 measured at room temperature, for three different values of the offset angle $\theta$ between the free and pinned-layer magnetic moments.

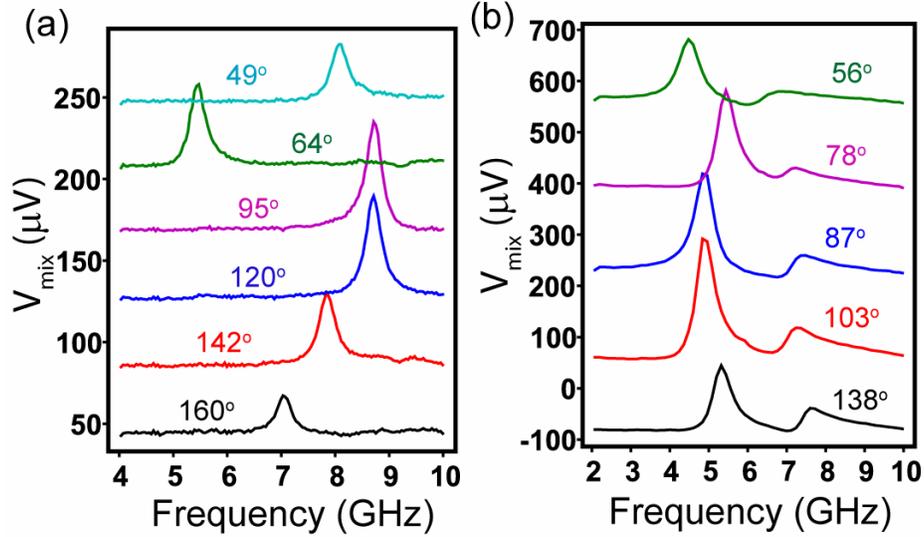

**Fig. 2**. (Color online) Selected ST-FMR resonance curves, measured at room temperature for selected initial offset angles $\theta$, for (a) sample #1 and (b) sample #3. The data are artificially offset in the vertical direction. The incident power in (a) is $P_{inc} = 2.7$ µW, and in (b) it is $P_{inc} = 5.0$ µW. The magnitude and angle (measured relative to the exchange bias direction) of the applied magnetic field corresponding to the various values of $\theta$ are (a) $\theta = 49°$: $H = 700$ Oe at 70°, $\theta = 64°$: $H = 400$ Oe at 70°, $\theta = 95°$: $H = 800$ Oe at 120°, $\theta = 120°$: $H = 700$ Oe at 140°, $\theta = 142°$: $H = 500$ Oe at 150°, and $\theta = 160°$: $H = 300$ Oe at 160°, (b) $\theta = 56°$: $H = 200$ Oe at 62°, $\theta = 78°$: $H = 200$ Oe at 88°, $\theta = 87°$: $H = 200$ Oe at 92°, $\theta = 103°$: $H = 200$ Oe at 118°, $\theta = 138°$: $H = 200$ Oe at 147°.